\documentclass[twocolumn,secnumarabic,amssymb,nobibnotes,aps,prd,showpacs,showkeys]{revtex4-1}

\usepackage{graphicx}
\usepackage{amsmath}
\usepackage{hyperref}
\usepackage{bbold}
\usepackage{footmisc}

\hypersetup{
	colorlinks=true,
	citecolor=blue,
	linkcolor=blue,
	filecolor=blue,      
	urlcolor=blue,
}

\newcommand{\aof}{\mathring{a}_{\text{of}}^{(3)}}

\makeatletter
\def\@fnsymbol#1{\ensuremath{\ifcase#1\or \dagger\or *\or \ddagger\or
		\mathsection\or \mathparagraph\or \|\or **\or \dagger\dagger
		\or \ddagger\ddagger \else\@ctrerr\fi}}
\makeatother

\begin{document}
	
	\title{Probing Lorentz violation in $2\nu\beta\beta$ using single electron spectra and angular correlations} 
	
	\author{ O.V. Ni\c{t}escu$^{1,2,3}$, S.A. Ghinescu$^{1,2,3}$, M. Mirea$^{1,2}$}
	\thanks{Deceased, 28.08.2020}
	
	\author{S. Stoica$^{1,2}$}
	\email[Corresponding author:]{sabin.stoica@cifra-c2unesco.ro}
	\affiliation{
		$^1$ "Horia Hulubei" National Institute of Physics and Nuclear Engineering, 30 Reactorului, POB MG-6, RO-077125, Bucharest-M\u agurele, Romania \\
		$^2$International Centre for Advanced Training and Research in Physics, PO Box MG12, 077125-Magurele, Romania\\
		$^3$ Faculty of Physics, University of Bucharest, 405 Atomi\c stilor, POB MG-11, RO-077125, Bucharest-M\u agurele, Romania}
	
	\date{\today}
	
	\begin{abstract}		
		We show that the current search for Lorentz invariance violation (LIV) in the summed energy spectra of electrons in $2\nu\beta\beta$ decay can be extended by investigating the single electron spectra and the angular correlation between the emitted electrons. We derive and calculate the LIV contributions to these spectra associated with the anisotropic part of the countershaded operator and controlled through the coefficient $\aof$ and discuss possible signatures that may be probed in experiments. First, we show that some distortion occurs in the single electron spectrum, maximal at small electron energies.  Then, we show that other LIV effects may be highlighted by analysing the angular correlation spectra and the ratio between the Standard Model Extension (SME) electron spectra and their Standard Model (SM) forms. We found that these LIV signatures depend on the magnitude of $\aof$, manifest differently for positive and negative values of this coefficient, and become more pronounced as the electron energy approaches the $Q$-value. Finally, we propose an alternative, new method to constrain $\aof$ through the measurement of the angular correlation coefficient. Using this method, and considering only statistical uncertainties, we obtain bounds of $\aof$ at the level of present ones, obtained from summed energy spectra. We show that future experiments can improve these limits significantly. Our study is performed for $^{100}$Mo, but the results hold qualitatively for other nuclei that undergo a double-beta decay. We hope our results will provide additional motivation for the LIV analyses performed in DBD experiments.
	\end{abstract}
	
	
	
	\maketitle
	
	\setcounter{equation}{0}
	\renewcommand{\theequation}{\arabic{equation}}
	
	
	\paragraph*{Introduction.}Searching for evidence to probe the Lorentz invariance violation (LIV) is a very current topic that joins the increasing effort to test the limits of the Standard Model (SM) \cite{KAS-PRD39,KP-PRD51}. The theoretical basis of these searches is the SM extension (SME), an effective field theory including operators that break Lorentz invariance for all the particles in the SM \cite{KP-PRD51,KOS-PRD69}.
	In particular, the neutrino sector of SME provides the theoretical framework for a rich phenomenology for searching evidence of LIV, 
	for example, those that can be proved in neutrino oscillations experiments \cite{KM-PRD69,KM-PRD70,DKM-PRD80, Diaz-AHEP}. However, there are LIV signatures related to the so-called countershaded effects associated with the oscillation-free operators of mass dimension three, which cannot be investigated in such experiments. The study of beta and double-beta decays offers the possibility to investigate the LIV effects related to the time-like (isotropic) component of this oscillation-free operator whose size is controlled by the coefficient $\aof$. In refs.  \cite{Diaz-PRD88, Diaz-PRD89} the LIV effects in $2\nu\beta\beta$ decay were calculated for the summed energy spectra of electrons, but employing a non-relativistic approximation for the electron radial wave functions. At present, the accuracy required by the DBD experiments far exceeds this approximation.  Recently, experiments like EXO-200 \cite{EXO-200-PRD93}, CUPID-0 \cite{CUPID-0-PRD100}, NEMO-3 \cite{NEMO-3-2019}, CUORE \cite{CUORE-2019, CUORE-PhDThesis}, GERDA \cite{GERDA-PhDThesis} have provided limits of the $\aof$ parameter through a careful analysis of the summed energy spectra of electrons in $2\nu\beta\beta$ decays, using theoretical spectra obtained with better but still approximate methods of calculation. In a recent paper \cite{NIT-2020}, we examined the effects of LIV on summed energy spectra of electrons and quantities related to them using Fermi functions built with exact electron wave functions. These were obtained by numerically solving the Dirac equation in a realistic Coulomb-type potential including the finite nuclear size, diffuse nuclear surface, and screening effects \cite{SM-2013, MPS-2015}. 

	In this work, we show that LIV signatures may also be searched in the single electron spectra and the angular correlation between the two electrons emitted in $2\nu\beta\beta$ decay. First, we derive the LIV contributions to these spectra and calculate them using an improved version of our method described in refs. \cite{SM-2013, NIT-2020}. Then, we discuss possible signatures that could be probed in experiments. We show that some distortion occurs in the single electron spectrum, maximal at small electron energy. Then, we present other possible LIV effects that can be highlighted by analysing the angular correlation spectra and the ratio between the total electron spectra (SME) and their standard counterparts (SM). We found that these LIV contributions manifest differently for positive and negative values of the $\aof$ coefficient, increasing in magnitude as the electron energy gets close to the $Q$-value. Finally, we propose an alternative, new method for constraining the $\aof$ coefficient, without a detailed analysis of the electron spectra, namely through the measurement of the angular correlation coefficient, $k^{2\nu}$. With a rough estimation, considering only the statistical errors, we constrained $\aof$ at the level of the present bounds obtained from summed energy electron spectra. We show that future experiments capable of measuring the angular correlation between electrons can significantly improve these limits. Our study is performed for the nucleus $^{100}$Mo, but the results hold qualitatively for other nuclei that undergo a double-beta decay.

	\paragraph*{Formalism.}The differential decay rate for the standard $2\nu\beta\beta$ process, $0^+\rightarrow0^+_1$ transitions, can be expressed as  \cite{Haxton-1984, Doi-1985, Haxton-1984,Tomoda-1991,Kotila-2012}
	\begin{equation}
	d\Gamma^{2\nu}=\left[\mathcal{A}^{2\nu}+ \mathcal{B}^{2\nu}\cos\theta_{12}\right]w^{2\nu}d\omega_1d\varepsilon_1d\varepsilon_2d(\cos\theta_{12}),
	\label{eq:DiferentialRate}
	\end{equation}
	where $\varepsilon_{1,2}$ are the electron energies, $\omega_{1,2}$ are the anti-neutrino energies, and $\theta_{12}$ is the angle between the two emitted electrons. In what follows, we adopt the natural units ($\hbar=c=1$). Within the SM framework the $w^{2\nu}$ quantity reads 
	\begin{equation}
	w^{2\nu}_{\text{SM}}=\frac{g_A^4G_F^4\left|V_{ud}\right|^4}{64\pi^7}\omega_1^2\omega_2^2p_1p_2\varepsilon_1\varepsilon_2,
	\end{equation} 
	where $g_A$ is the axial vector constant, $G_F$ is the Fermi coupling constant, $V_{ud}$ is the first element of the Cabibbo-Kobayashi-Maskawa matrix and $p_{1,2}$ are the momenta of the electrons.
	
	The quantities $\mathcal{A}^{2\nu}$ and $\mathcal{B}^{2\nu}$ can be expressed to a good approximation by \cite{Tomoda-1991}
	\begin{align}
	\begin{aligned}
	\label{eq:NMESeparation}
	\mathcal{A}^{2\nu}=&\frac{1}{4}a(\varepsilon_1,\varepsilon_2)\left|M_{2\nu}\right|^2\tilde{A}^2\times\\&\left[\left(\langle K_N\rangle+\langle L_N\rangle\right)^2+
	\frac{1}{3}\left(\langle K_N\rangle-\langle L_N\rangle\right)^2\right],\\
	\mathcal{B}^{2\nu}=&\frac{1}{4}b(\varepsilon_1,\varepsilon_2)\left|M_{2\nu}\right|^2\tilde{A}^2\times\\&\left[\left(\langle K_N\rangle+\langle L_N\rangle\right)^2-
	\frac{1}{9}\left(\langle K_N\rangle-\langle L_N\rangle\right)^2\right],
	\end{aligned}
	\end{align}
	\noindent
	where $M_{2\nu}$ are the nuclear matrix elements (NMEs) and $a(\varepsilon_1,\varepsilon_2)$ and $b(\varepsilon_1,\varepsilon_2)$ are products of the radial wave functions of the emitted electrons. $\langle K_N\rangle$, $\langle L_N\rangle$ are kinematic factors that depend on the electron and anti-neutrino energies, on the ground state energy $E_I$ of 
	the parent nucleus and on an averaged energy $\langle E_N \rangle$ of the excited $1^+$ states in the intermediate nucleus (closure approximation). The expressions for the kinematic factors are given by \cite{Haxton-1984}
	\begin{align}
	\begin{aligned}
	\label{eq:KnDef}
	\langle K_N\rangle=
	{1\over \varepsilon_1+\omega_1+\langle E_N\rangle-E_I}+
	{1\over \varepsilon_2+\omega_2+\langle E_N\rangle-E_I}\\
	\langle L_N\rangle=
	{1\over \varepsilon_1+\omega_2+\langle E_N\rangle-E_I}+
	{1\over \varepsilon_2+\omega_1+\langle E_N\rangle-E_I}.
	\end{aligned}
	\end{align}
	Here, the difference in energy in the denominator can be obtained from the approximation
	$\tilde{A}^2=[W_0/2+\langle E_N\rangle -E_I]^2$, where
	$\tilde{A}=1.12A^{1/2}$ (in MeV) gives the energy of the giant Gamow-Teller resonance in the intermediate nucleus. The energy $W_0$ is defined as
	\begin{equation}
	W_0=Q+2m_e=E_I-E_F,
	\end{equation}
	where $ Q $ is the kinetic energy available for the four leptons, $m_e$ is the rest energy of the electron, and $ E_F $ is the ground state energy of 
	the final nucleus.
	
	The functions $a(\varepsilon_1,\varepsilon_2)$ and $b(\varepsilon_1,\varepsilon_2)$ are defined as
	\begin{align}
	\begin{aligned}
	&a(\varepsilon_1,\varepsilon_2)=\left|\alpha^{-1-1}\right|^2+\left|\alpha_{11}\right|^2+\left|\alpha_{1}^{\hspace{0.16cm}-1}\right|^2+\left|\alpha^{-1}_{\hspace{0.35cm}1}\right|^2,\\
	&b(\varepsilon_1,\varepsilon_2)=-2\Re\{\alpha^{-1-1}\alpha_{11}^*+\alpha^{-1}_{\hspace{0.35cm}1}\alpha_{1}^{\hspace{0.16cm}-1*}\},
	\end{aligned}
	\end{align}
	with
	\begin{align}
	\label{eq:WavefunctionsProducts}
	\begin{aligned}
	\alpha^{-1-1}&=g_{-1}(\varepsilon_1)g_{-1}(\varepsilon_2),
	\alpha_{11} = f_{1}(\varepsilon_1)f_{1}(\varepsilon_2),\\
	\alpha_{1}^{\hspace{0.16cm}-1}&=f_{1}(\varepsilon_1)g_{-1}(\varepsilon_2),
	\alpha^{-1}_{\hspace{0.35cm}1}= g_{-1}(\varepsilon_1)f_{1}(\varepsilon_2).
	\end{aligned}
	\end{align}
	\noindent
	The functions $f_{1}(\varepsilon_1)$ and $g_{-1}(\varepsilon_2)$ are the electron radial wave functions evaluated on the surface of the daughter nucleus as
	\begin{align}
	\begin{aligned}
	g_{-1}(\varepsilon)&=\int_{0}^{\infty}g_{-1}(\varepsilon,r)\delta(r-R)dr,\\
	f_{1}(\varepsilon)&=\int_{0}^{\infty}f_{1}(\varepsilon,r)\delta(r-R)dr,
	\end{aligned}
	\end{align}
	where $R=r_0A^{1/3}$, $r_0=1.2$ fm. 
	
	The derivation of the decay rate with respect to the cosine of the angle $\theta_{12}$ can be expressed as \cite{Doi-1985}
	\begin{equation}
	\frac{d\Gamma^{2\nu}_\text{SM}}{d(\cos\theta_{12})}=\frac{1}{2}\Gamma^{2\nu}_{\text{SM}}\left[1+\kappa^{2\nu}_{\text{SM}} \cos\theta_{12}\right],
	\end{equation}
	where $\kappa^{2\nu}_{\text{SM}}$ is the angular correlation coefficient defined by
	\begin{equation}
	\kappa^{2\nu}_{\text{SM}}=\frac{\Lambda^{2\nu}_{\text{SM}}}{\Gamma^{2\nu}_{\text{SM}}}.
	\end{equation}    
	
	The decay rates $\Gamma^{2\nu}_{\text{SM}}$ and $\Lambda^{2\nu}_{\text{SM}}$ are obtained by integrating Eq.~\eqref{eq:DiferentialRate} over the lepton energies. In the closure approximation their formulas can be written in a factorized form as follows
	\begin{align}
	\begin{aligned}
	\frac{\Gamma^{2\nu}_{\text{SM}}}{\ln 2}&=g_A^4\left|m_eM_{2\nu}\right|^2G^{2\nu}_{\text{SM}},\\
	\frac{\Lambda^{2\nu}_{\text{SM}}}{\ln 2}&=g_A^4\left|m_eM_{2\nu}\right|^2H^{2\nu}_{\text{SM}},
	\end{aligned}
	\end{align}
	which essentially are products of NMEs and phase space factors (PSFs) $G^{2\nu}_{\text{SM}}$ and $H^{2\nu}_{\text{SM}}$.
	
	Within the SME, the LIV effects in the neutrino sector can also arise from the action of the so-called countershaded operators, by changing each antineutrino 4-momentum from $q^{\alpha}=(\omega,\boldsymbol{q})$ to an effective 4-momentum
	$\tilde{q}^{\alpha}=(\omega,\boldsymbol{q}+\boldsymbol{a}_{\text{of}}^{(3)}-\mathring{a}_{\text{of}}^{(3)}\hat{\boldsymbol{q}})$ \cite{KR-RMP2011,Diaz-PRD89,KT-PRL102}. Here, $\aof$ is related to the isotropic component of the $(a_{\text{of}}^{(3)})_{jm}$ coefficient by $\aof = (a_{\text{of}}^{(3)})_{00}/\sqrt{4\pi}$ \cite{KR-RMP2011}. Since in  $2\nu\beta\beta$ experiments the two antineutrinos are not measured, the integration over all neutrino orientations leaves only the isotropic coefficient $\mathring{a}_{\text{of}}^{(3)}$, hence LIV effects related only to this contribution can be searched. This leads to a change in the form of the antineutrino differential phase space, from the standard one $d^3q=4\pi\omega^2d\omega$ to the one containing the LIV effects $d^3q=4\pi(\omega^2+2\omega\mathring{a}_{\text{of}}^{(3)})d\omega$. To the first order in $\aof$, the term $w^{2\nu}$ in the differential decay rate, i.e. Eq.~\eqref{eq:DiferentialRate}, acquires the form    
	\begin{align}
	\begin{aligned}
	\label{eq:LVPhaseSpaceTerm}
	w^{2\nu}_{\text{SME}}=&\frac{g_A^4G_F^4\left|V_{ud}\right|^4}{64\pi^7}p_1p_2\varepsilon_1\varepsilon_2\times\\&\left[\omega_1^2\omega_2^2+2\mathring{a}_{\text{of}}^{(3)} (\omega_1^2\omega_2+\omega_1\omega_2^2)\right].
	\end{aligned}
	\end{align}
	In the above expression the first term represents the SM contribution and the following two terms are the LIV contributions. Following the same steps as in the case of SM, we get the SME expression of the differential decay rate with respect to the cosine of the angle $\theta_{12}$
	\begin{equation}
	\frac{d\Gamma^{2\nu}_\text{SME}}{d(\cos\theta_{12})}=\frac{1}{2}\Gamma^{2\nu}_{\text{SME}}\left[1+\kappa^{2\nu}_{\text{SME}}\cos\theta_{12}\right],
	\end{equation}
	where the angular correlation coefficient $\kappa^{2\nu}_{\text{SME}}$ is defined by
	\begin{equation}
	\kappa^{2\nu}_{\text{SME}}=\frac{\Lambda^{2\nu}_{\text{SME}}}{\Gamma^{2\nu}_{\text{SME}}}.
	\end{equation}    
	The decay rates $\Lambda^{2\nu}_{\text{SME}}$ and $\Gamma^{2\nu}_{\text{SME}}$  can be expressed as sum of standard and LIV contributions 
	\begin{align}
	\begin{aligned}
	\Gamma^{2\nu}_{\text{SME}}&=\Gamma^{2\nu}_{00}+\Gamma^{2\nu}_{01}+\Gamma^{2\nu}_{10},\\
	\Lambda^{2\nu}_{\text{SME}}&=\Lambda^{2\nu}_{00}+\Lambda^{2\nu}_{01}+\Lambda^{2\nu}_{10},
	\end{aligned}
	\end{align} 
	where ${00}$ stands for the SM contribution. 
	The components of the decay rates can be also written in a good approximation in a factorized form as products NMEs and PSFs, as follows
	\begin{align}
	\label{eq:DecayRateSeparation}
	\begin{aligned}
	\frac{\Gamma^{2\nu}_{mn}}{\ln 2}&=g_A^4\left|m_eM_{2\nu}\right|^2G^{2\nu}_{mn},\\
	\frac{\Lambda^{2\nu}_{mn}}{\ln 2}&=g_A^4\left|m_eM_{2\nu}\right|^2H^{2\nu}_{mn},
	\end{aligned}
	\end{align}
	with $mn=\{00,10,01\}$ and $M_{2\nu}$ the NMEs. The PSF expressions can be written in the following compact form
	\begin{align}
	\begin{aligned}
	\label{eq:Phase-Space-Factors}
	\begin{Bmatrix}
	G^{2\nu}_{mn}\\
	H^{2\nu}_{mn}
	\end{Bmatrix}&=(10\mathring{a}_{\text{of}}^{(3)})^{m+n}\frac{C_{mn}}{m_e^{11-m-n}}
	\times\\
	&\int_{m_e}^{E_I-E_F-m_e}d\varepsilon_1\varepsilon_1p_1\int_{m_e}^{E_I-E_F-\varepsilon_1}d\varepsilon_2\varepsilon_2p_2\times\\
	&\int_{0}^{E_I-E_F-\varepsilon_1-\varepsilon_2}d\omega_1\Omega_{mn}\times\\
	&\hspace{-1cm}\begin{Bmatrix}
	a(\varepsilon_1,\varepsilon_2)\left(\langle K_N\rangle^2+\langle L_N\rangle^2+
	\langle K_N\rangle\langle L_N\rangle\right)\\
	b(\varepsilon_1,\varepsilon_2)\left[\frac{2}{3}\left(\langle K_N\rangle^2+\langle L_N\rangle^2\right)+\frac{5}{3}
	\langle K_N\rangle\langle L_N\rangle\right]
	\end{Bmatrix},
	\end{aligned}
	\end{align}
	with $\Omega_{mn}=\omega_1^{2-m}\omega_2^{2-n}$ and  
	\begin{align}
	\begin{aligned}
	C_{00}=&\frac{\tilde{A}^2 G_F^4 |V_{ud}|^4 m_e^9}{96\pi^7\ln 2}, \\
	C_{10}=C_{01}=&\frac{\tilde{A}^2 G_F^4 |V_{ud}|^4 m_e^8}{480\pi^7\ln 2}. 
	\end{aligned}
	\end{align}
	In this study, we consider that LIV influences only the PSFs. In the PSF definitions, the LIV parameter $\mathring{a}_{\text{of}}^{(3)}$ is included in MeV and the energy $\omega_2$ of the antineutrino is determined as $\omega_2=E_I-E_F-\varepsilon_1-\varepsilon_2-\omega_1$. We observe that the first order LIV contributions $(10)$ and $(01)$ are functions of $\omega_1$ symmetric to the center of the integration interval $[0,E_I-E_F-\varepsilon_1-\varepsilon_2]$, and hence they are equal in value. So, in what follows, we consider the corrections to the standard PSFs as
	\begin{equation}
	\begin{Bmatrix}
	\delta G^{2\nu}\\
	\delta H^{2\nu}
	\end{Bmatrix}=
	\begin{Bmatrix}
	\frac{2G^{2\nu}_{10}}{\mathring{a}_{\text{of}}^{(3)}}\\
	\frac{2H^{2\nu}_{10}}{\mathring{a}_{\text{of}}^{(3)}}
	\end{Bmatrix}.
	\end{equation} 
	We note that by making the approximation 
	\begin{equation}
	\label{eq:KnLnApprox}
	\langle K_N\rangle\simeq \langle L_N\rangle\simeq\frac{2}{E_I-\langle E_N\rangle-W_0/2},
	\end{equation}
	and integrating over the energy of the antineutrino $\omega_1$, one retrieves simplified expressions of the PSFs which were used in many previous works (see for example \cite{Suhonen-1998} and references therein) and also in the previous LIV analyzes \cite{EXO-200-PRD93,CUPID-0-PRD100,NEMO-3-2019}.
	Deriving the decay rate expression versus the kinetic energy of one electron and to the total kinetic energy of the two electrons we get the single electron spectrum
	\begin{equation}\label{eq:SingleElectronSpectra}
	\frac{d\Gamma^{2\nu}_{\textrm{SME}}}{d\varepsilon_1} = C\frac{dG^{2\nu}_{00}}{d \varepsilon_1}\left(1+\aof \chi^{(1)}(\varepsilon_1)\right),
	\end{equation}
	and the summed energy spectrum of electrons
	\begin{equation}\label{eq:SumElectronSpectra}
	\frac{d\Gamma^{2\nu}_{\textrm{SME}}}{dK} = C\frac{dG^{2\nu}_{00}}{d K}\left(1+\aof \chi^{(+)}(K)\right),
	\end{equation}
	where $C$ is a constant including NME, $ K\equiv \varepsilon_1 + \varepsilon_2 -2m_e $ is the total kinetic energy of the two electrons and 
	\begin{align}\label{eq:LIVdeviations}
	\begin{aligned}
	\chi^{(1)}(\varepsilon_1) = \frac{d(\delta G^{2\nu})}{d \varepsilon_1}/\frac{dG^{2\nu}_{00}}{d \varepsilon_1}, \\
	\chi^{(+)}(K) = \frac{d(\delta G^{2\nu})}{d K}/\frac{dG^{2\nu}_{00}}{d K}, 
	\end{aligned}
	\end{align}
	incorporate the deviations of the electron spectra from their SM forms. 
	Deriving also the decay rate versus $\varepsilon_1$ and $\cos(\theta_{12})$, we get the expressions of the angular correlation and its deviation from the SM form due to LIV
	\begin{align}
	\label{eq:DiffDecayRate_SME}
	\begin{aligned}
	&\frac{d\Gamma^{2\nu}_\text{SME}}{d \varepsilon_1 d(\cos\theta_{12})}=C\frac{d G^{2\nu}_{00}}{d\varepsilon_1}\times\\
	&\left[1+\aof \chi^{(1)}(\epsilon_1)+\left(\alpha^{2\nu}_{\text{SM}}+\aof\frac{d(\delta H^{2\nu})/d\varepsilon_1}{dG^{2\nu}_{00}/d\varepsilon_1}\right)\cos\theta_{12}\right]
	\end{aligned}
	\end{align}
	where $ \alpha_{\text{SM}} \equiv (dH^{2\nu}_{00}/d\varepsilon_1)/(dG^{2\nu}_{00}/d\varepsilon_1)$ is the SM angular correlation while its SME expression is
	\begin{equation}
	\label{eq:alpha_sme}
	\alpha_{\text{SME}} = \alpha_{\text{SM}} + \aof \frac{d(\delta H^{2\nu})/d\varepsilon_1}{dG_{00}^{2\nu}/d\varepsilon_1}.
	\end{equation}
	Deriving the decay rate expression versus $\cos(\theta_{12})$
	\begin{align}
	\label{eq:k_sme}
	\begin{aligned}
	&\frac{d\Gamma^{2\nu}_\text{SME}}{d(\cos\theta_{12})}=CG^{2\nu}_{00}\times\\
	&\left[1+\aof\frac{\delta G^{2\nu}}{G^{2\nu}_{00}}+\left(\kappa^{2\nu}_{\text{SM}}+\aof\frac{\delta H^{2\nu}}{G^{2\nu}_{00}}\right)\cos\theta_{12}\right],
	\end{aligned}
	\end{align}
	we can identify (in round brackets) the SME expression of the angular correlation coefficient $\kappa^{2\nu}_{\text{SME}}$. For an independent treatment with respect to $\aof$, we define $\xi_{LV}^{2\nu} \equiv \delta H^{2\nu}/G^{2\nu}_{00}$ in units of $\mathrm{MeV}^{-1}$. 
	\paragraph*{Results and discussion.} We apply the formalism deduced in the previous section to the $^{100}$Mo nucleus. For this isotope, there is strong experimental evidence for the single state dominance (SSD) hypothesis \cite{Simkovic_2001,Domin-2005}, namely that only the first $1^+$ state in the intermediate odd-odd nucleus contribute to the decay \cite{NEMO-3-2019}. Consequently, we adopt the SSD hypothesis in our calculations, which amounts to replacing the averaged energy, $ \langle E_N\rangle $, with the energy of the dominant state, $ E_{1^+} $ = $-0.343 \mathrm{MeV}$ \cite{Kotila-2012,Simkovic_2001}. Then, we used our method for the PSF computation, which is presented in detail in refs. \cite{SM-2013, MPS-2015}. To obtain the exact electronic radial wave functions, we improved our numerical procedure using the RADIAL package \cite{Salvat-1991, Salvat-1995}.  
	\begin{figure}[ht!]
		\includegraphics[width=0.49\textwidth]{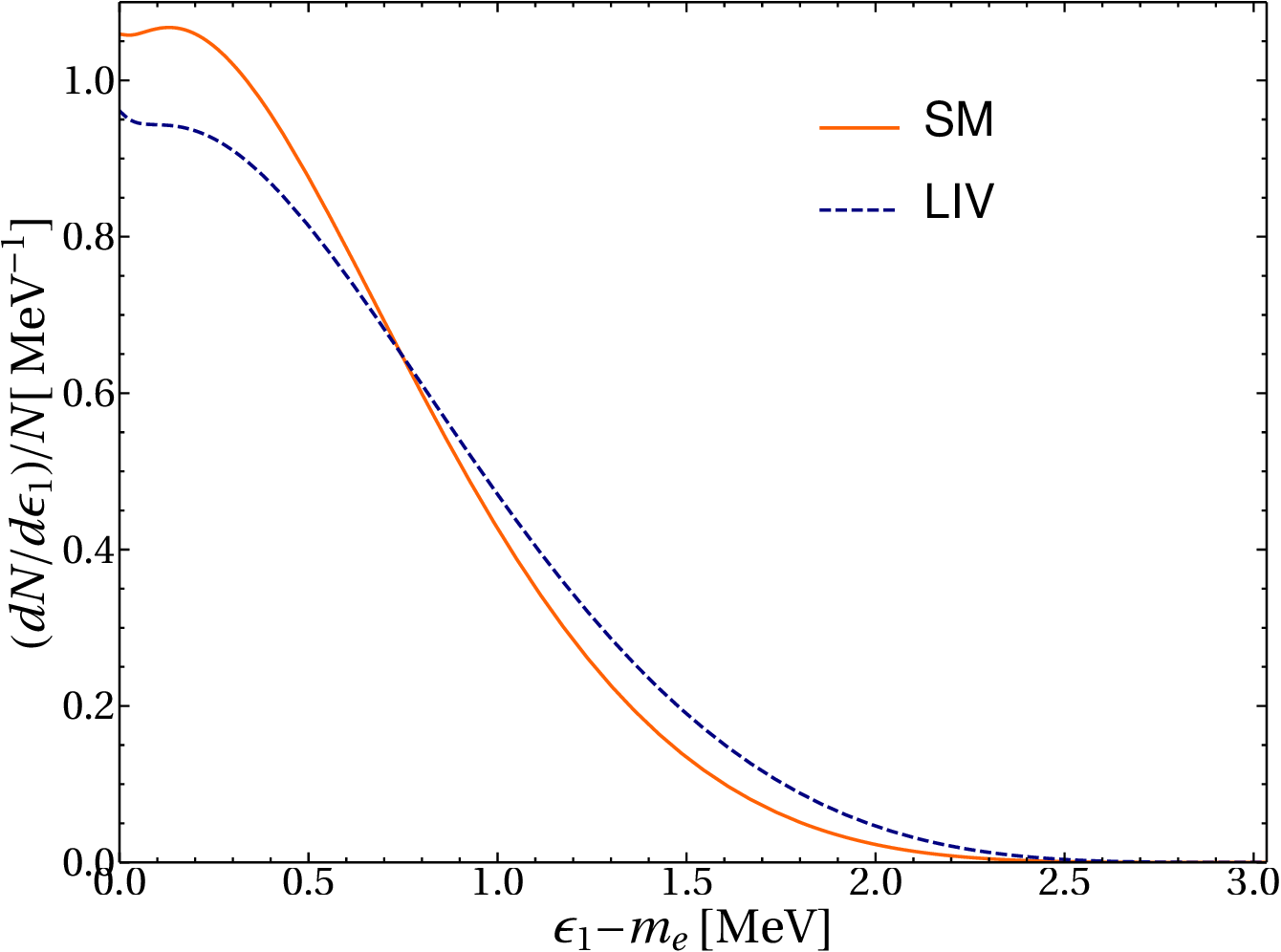}
		\caption{(Color online) Normalized $2\nu\beta\beta$ single electron spectra within SM with solid line and the first order contribution in $\aof$ due to LIV with dashed line, for $ ^{100} $Mo in the SSD hypothesis.  }
		\label{fig:SingleElectronSpectra}
	\end{figure}
	
	The numerical results are obtained using the following physical constants: the electron mass $ m_e = 0.5110 \mathrm{MeV} $, the CKM matrix element $ V_{ud} =  0.9743$, the Fermi coupling constant $ G_F =  1.1666 \times 10^{-11} \mathrm{MeV}^{-2}$, the fine structure constant $ \alpha = 1/137.036 $ \cite{PDG-statistics} and the Q-value of $ ^{100}\mathrm{Mo} $ $ 2\nu\beta\beta $ decay $ Q = 3.0344 \mathrm{MeV}$ \cite{Ge-PLB662-2008}. In the calculations we use two sets of $\aof$ limits, the one reported by the EXO-200 collaboration ($-2.65\times 10^{-2}\mathrm{MeV} \le \aof \le 7.6\times 10^{-3}\mathrm{MeV}$) \cite{EXO-200-PRD93} and the one reported by the NEMO-3 collaboration ($-4.2\times 10^{-4}\mathrm{MeV} \le \aof \le 3.5\times 10^{-4}\mathrm{MeV}$) \cite{NEMO-3-2019}, as they represent the least and most stringent limits, respectively, reported until now. 	
	
	In Fig.\ref{fig:SingleElectronSpectra} we illustrate the normalized single electron energy spectrum and its deviation due to LIV. Our predictions indicate some distortion of the SM spectrum, maximal at very small electron energy, unlike a similar effect that was predicted in the summed energy electron spectra \cite{EXO-200-PRD93, NIT-2020}, and which is maximal at larger electron energy. We note that the maximum at very small electron energy is due to the use in calculation of the SSD prescription, namely for the nuclei where SSD  is valid, a similar distortion of the single electron spectra at small energies, due to LIV, occurs as well. The calculation shows that this LIV effect is larger in the single electron spectra than in the summed energy ones, but the statistics for its observation is smaller than in the last case. However, it is worth to mention that such LIV effects in the electron spectra can be presently detected only as global distortions of these spectra. With higher statistics planned for future DBD experiments, such perturbations due to LIV might become detectable, and accurate theoretical predictions of the electron spectra and angular correlation are essential to interpret the data and constrain the $\aof$ coefficient. Our results indicate that both single and summed energy spectra of electrons can investigated to detect similar distortions due to LIV.   
	
	\begin{figure}[ht!]	
		\includegraphics[width=0.49\textwidth]{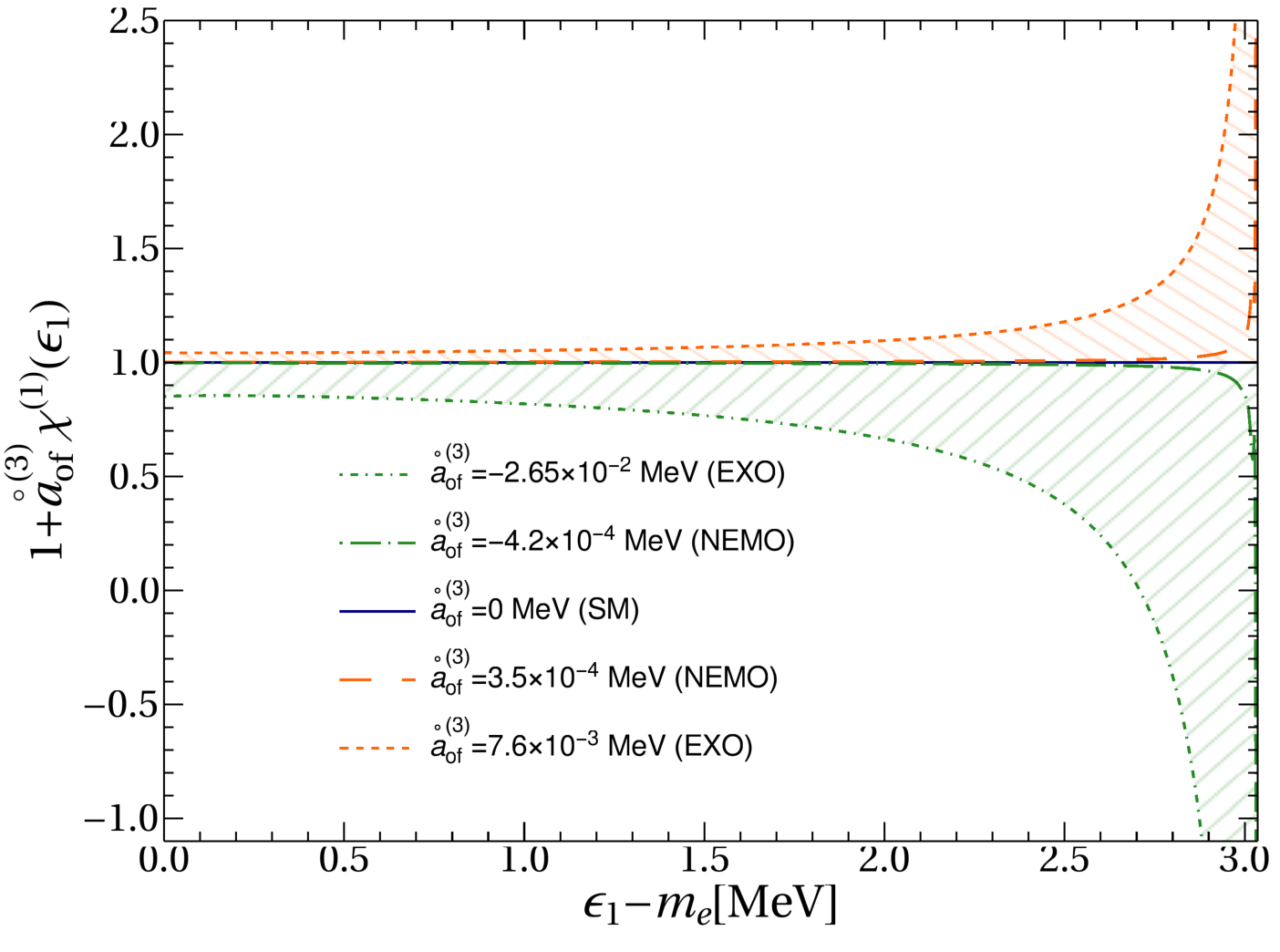}
		\caption{(Color online) The quantity $1+\aof\chi^{(1)}(\varepsilon_1)$ depicted for current limits of $\aof$ (dashed for upper limit and dot-dashed for lower limit). The solid line represents the SM prediction.  }
		\label{fig:ChiSingle}
	\end{figure} 
	
	Then, we analyse the ratio between the total single electron spectrum (including LIV contribution) and its SM form, namely $1 + \aof \chi^{(1)}(\varepsilon_1)$. We plot this quantity in Fig.~\ref{fig:ChiSingle} where other LIV effects are observed. With solid line we plotted the standard (SM) spectrum, with dashed line the total spectrum for positive values of $\aof$ and with dot-dashed line the total spectrum for negative values of $\aof$. One observes that the curves are very close to each other, but they diverge at electron energies approaching $ Q $-value. This divergence is due to a slower descent (in absolute value) of the spectrum with respect to the SM one at the end of the energy interval. This effect is quite visible for the $\aof$ limits reported by EXO-200, while for the more stringent ones reported by NEMO-3, it becomes practically un-observable. We note that searching for such LIV effect just in the vicinity of the $Q$-value is difficult, due to the current small experimental statistics but, as we already mentioned, looking for it as a global distortion of the ratio ($1+\aof\chi^{(1)}(\varepsilon_1)$), could be possible.  
	
\begin{figure}[ht!]
		\includegraphics[width=\columnwidth]{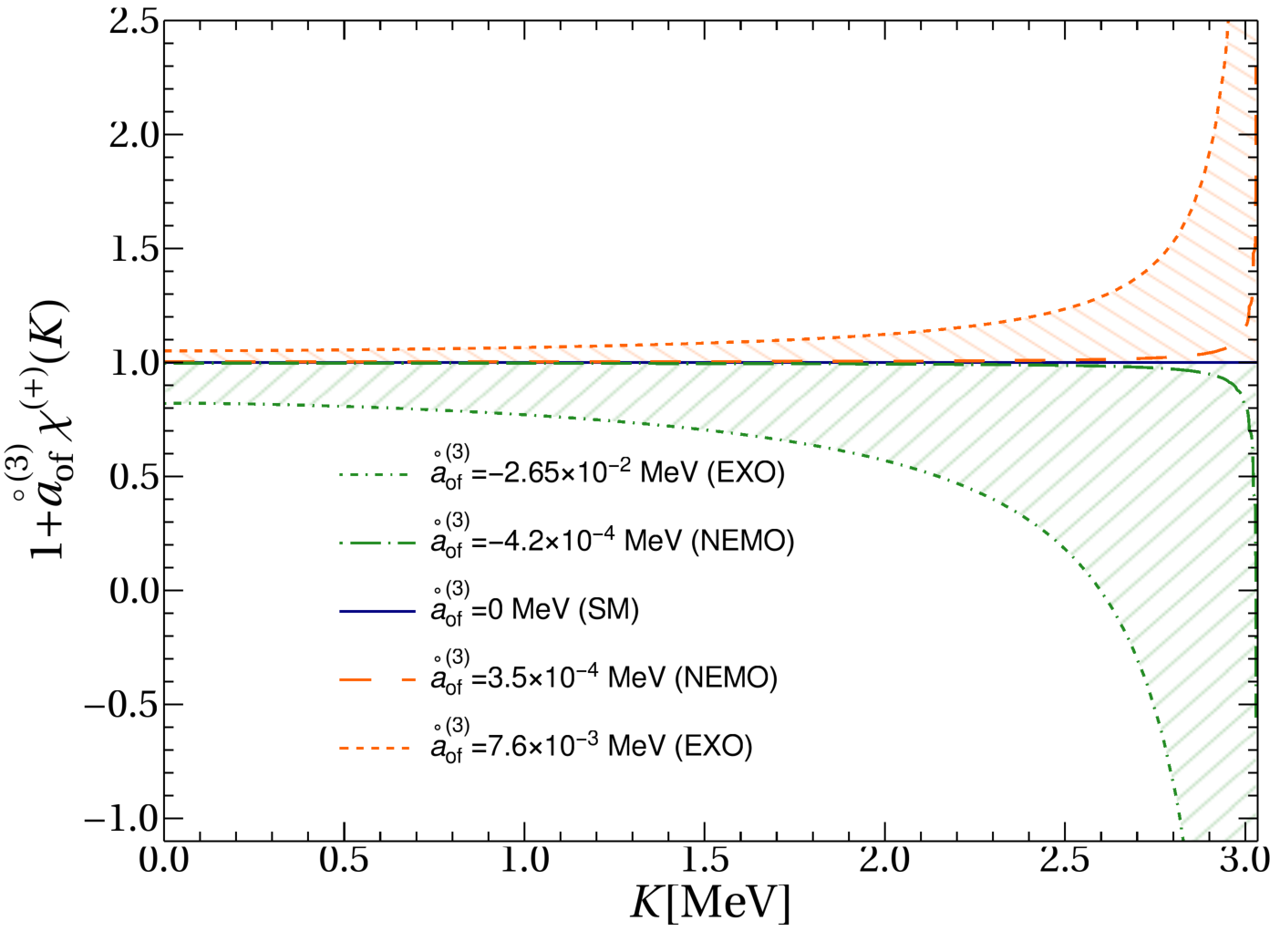}
		\caption{(Color online)The quantity $1+\aof\chi^{(+)}(K)$ depicted for current limits of $\aof$.The same conventions as in Fig.~\ref{fig:ChiSingle} are used.}
		\label{fig:ChiSummed}
	\end{figure}

	Next, we perform the same study for the summed energy spectrum of the electrons, plotting the quantity $ 1 + \aof\chi^{(+)}(K) $. As seen in Fig.~\ref{fig:ChiSummed} (we use the same notations as in Fig.~\ref{fig:ChiSingle}), an effect similar to that of the single electron spectrum also occurs and has the same explanation, namely the LIV contribution becomes higher than the standard spectrum itself at energies close to the $Q$-value. 
	It is also seen that for negative values of $ \aof $, the quantity $1+\aof \chi^{(+)}$ cut the $K$-axis at a certain summed energy of electrons, $K$. This feature is illustrated in Fig.~\ref{fig:ChiSummed} and is most visible only for $ \aof$ limits provided by EXO-200 collaboration. Beyond this $K$ value, the differential rate becomes negative, and thus the DBD process becomes forbidden. Conversely, for positive $\aof$, the decay rate is non-zero for some range $K > Q$. The general accepted interpretation of this behavior is a possible shift in the $Q$-value of the process \cite{Diaz-PRD88}. The same considerations also hold for the LIV effects described above, in single electron spectra.     

	The deviations from the standard spectra presented in Fig.~\ref{fig:ChiSingle} and Fig.~\ref{fig:ChiSummed} may be investigated experimentally by dividing the measured spectra by the theoretical prediction calculated within SM. The analysis will include the uncertainties for each energy bin measured. The best fit of the data can conclude if the quantity $1+\aof\chi^{(1)}(\varepsilon_1)$ or $1+\aof\chi^{(+)}(K)$ is increasing or decreasing, and so if the $\aof$  coefficient is positive or negative, respectively.    
	
	Further, we discuss the implications of the LIV contributions on the angular correlation spectrum, $\alpha^{2\nu}$, and the coefficient $k^{2\nu}_{\text{SME}}$. They appear in Eqs.~(\ref{eq:alpha_sme}) and (\ref{eq:k_sme}) as additional terms to the standard quantities, controlled by the $ \aof $ coefficient.
	
	\begin{figure}[ht!]
		\includegraphics[width=0.49\textwidth]{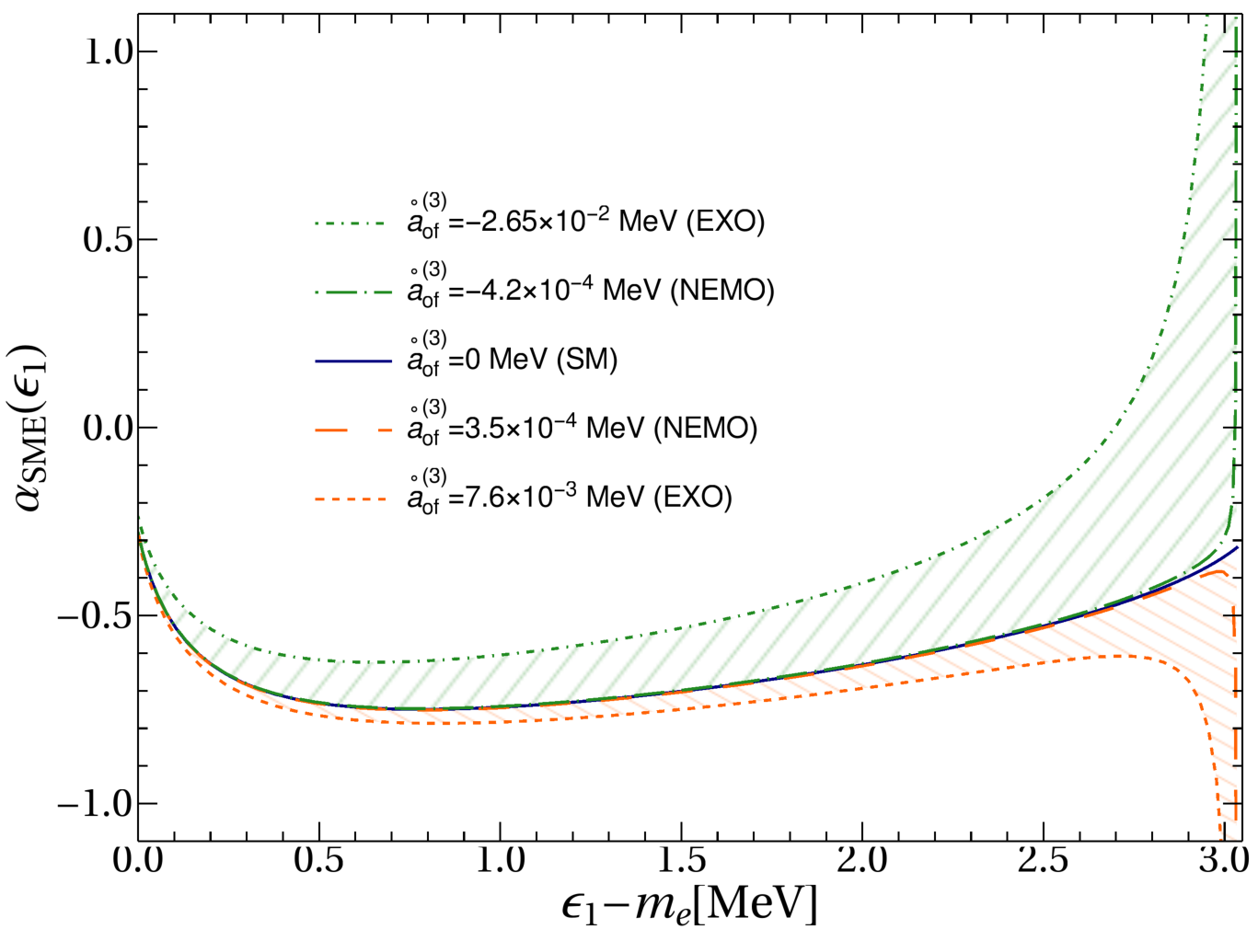}
		\caption{(Color online) The angular correlation spectrum plotted for the current limits of $\aof$. The same conventions as in Fig.~\ref{fig:ChiSingle} are used.}
		\label{fig:AngularCorrelationPositiveaof}
	\end{figure} 
	
	Fig.~\ref{fig:AngularCorrelationPositiveaof} depicts the angular correlation spectrum for the same sets of the $\aof$ limits. We note the total angular correlation spectrum for negative values of $\aof$ exceeds the SM spectrum since $\delta H$ is also negative, making the LIV contribution positive. As seen, there is similar, divergent behavior of the curves (even more pronounced than in the case of the electron spectra) as the electron energy increases, this time because $ \delta H^{2\nu} $ decreases slower than $ G_{00}^{2\nu} $. For the $\aof$ limits provided by EXO-200, this effect is quite large even over a large energy region, while using the NEMO-3 limits, the effect becomes negligible. 
	
	For completeness, we also calculate the LIV effect on the angular correlation coefficient, $  k^{2\nu} $. Its value is related to the PSF expressions from Eq.~(\ref{eq:Phase-Space-Factors}), which we computed using our method described in \cite{SM-2013, MPS-2015}, within the SSD hypothesis. First, we computed its SM value
	\begin{equation*}
	k_{\mathrm{SM}}^{2\nu} = -0.6676.
	\end{equation*}
	We note that this value differs from that reported in \cite{Simkovic-2020}, but the difference may stem from the fact that we used exact electronic wave functions instead of analytic, approximate ones, in the construction of the Fermi functions.  Using the same Fermi functions as in \cite{Simkovic-2020}, we obtained $k^{2\nu}_{\textrm{SM}}=-0.6365$, a value much closer to that reported in this reference. Then, we evaluate the LIV contribution
	\begin{equation*}
	\xi_{LV}^{2\nu} = -4.285 ~\mathrm{MeV}^{-1},
	\end{equation*}
	which can be plugged back into Eq.~(\ref{eq:k_sme}) to obtain  its SME value
	\begin{equation*}
	k_{\mathrm{SME}}^{2\nu} = -0.6676 -4.285\times \aof,
	\end{equation*}
	where $\aof$  must be taken in units of $ \mathrm{MeV} $. On the other hand, experimentally, the angular correlation coefficient can be determined via the forward-backward asymmetry defined by \cite{Arnold-2010}
	
	\begin{align}
	\mathcal{A}^{2\nu}\equiv&\left(\int^0_{-1}\frac{d\Gamma^{2\nu}}{dx}dx-\int^1_{0}\frac{d\Gamma^{2\nu}}{dx}dx\right)/\Gamma\nonumber\\
	=&\frac{N_{+}-N_{-}}{N_{+}+N_{-}}=\frac{1}{2}k_{\mathrm{SM}}^{2\nu},
	\end{align}
	where $x=\cos\theta_{12}$ and $N_{-}(N_{+})$ are the $2\nu\beta\beta$ events with the angle $\theta_{12}$ smaller (larger) than $\pi/2$. 
	For a number of $N=5\times10^{5}$ events at NEMO-3 \cite{NEMO-3-2019} and considering only the statistical errors, the angular correlation coefficient is measurable with the uncertainty $k_{\mathrm{SM}}^{2\nu}=0.6676\pm0.0027$. Without a statistically significant deviation from the SM expectation, we obtain a bound $|\aof|\lesssim 1.04\times 10^{-3}$ MeV at 90\% CL. This is only a rough estimation, and dedicated experimental analysis, including the systematic uncertainties, is necessary for a better one. We note that this estimation lies between the $\aof$ limits reported by NEMO-3 and EXO-200, which were obtained from the analysis of the summed energy spectra of electrons. We note here that if in a future experiment the number of $2\nu\beta\beta$ events would increase by three orders of magnitude (as planned for example in the SuperNEMO experiment), our estimation yields $|\aof|\lesssim 3.3\times 10^{-5}$ MeV at $90\%$ CL, which is comparable with the limits obtained from tritium decay experiments \cite{Diaz-PRD88}. Thus, we predict good perspectives for searching for LIV effects in future DBD experiments, due to the significant increase of statistics. 
	
	\paragraph*{Conclusions.} Concluding, we propose that the investigation of LIV effects in $2\nu\beta\beta$ to be extended to single electron spectra and the angular correlation between the two electrons. We derived the LIV contributions to the standard spectra and provided theoretical predictions for them to be used in the experimental analyses. Our study refers to the case of $^{100}\mathrm{Mo}$, but similar results are expected for other nuclei. We found some distortion in the single electron spectrum introduced by LIV (different from the distortion reported in the literature for the summed energy spectra). Then, we presented some other LIV signatures that may be experimentally investigated. For example, if the angular correlation spectrum and the ratios between the total electron spectra and their standard forms are analysed, distortions may occur. We show that they depend on the sign and magnitude of the $\aof$ coefficient and increase in magnitude as the electron energies approach the $Q$-value. Moreover, the LIV effects may shift the end-point of the allowed energies.    
	
	Finally, we propose an alternative, new method for constraining $\aof$, namely by analysing the LIV effects on the angular correlation coefficient. This coefficient is calculated from PSFs of the $2\nu\beta\beta$ decay and can be determined experimentally via the forward-backward asymmetry of emitted electrons. Using the present NEMO-3 sensitivity and taking into account only statistical errors, we obtained bounds for $\aof$ situated between the limits reported by NEMO-3 and EXO-200. Considering the expected performances of the future DBD experiments (high statistics, very low backgrounds, improved methods of measurement), we appreciate promising perspectives of these experiments to perform new relevant LIV investigations and to significantly improve the $\aof$ constrains. We hope that our study will contribute additional, useful information to motivate the LIV investigations in DBD. In this respect, we mention that we can provide, at request, the numerical data for the construction of the theoretical spectra necessary in the LIV analyses.
	
	\paragraph*{Acknowledgments.}	
	The authors wish to thank Jorge S. D\'{\i}az  and Viktor I. Tretyak  for the insightful discussions.
	The figures for this article have been created using the SciDraw scientific figure preparation system \cite{SciDraw}.
	This work has been supported by the grants of the Romanian Ministry of Education and Research through the projects UEFISCDI-18PCCDI/2018 and PN19-030102-INCDFM.

	
	\bibliography{thebibliography}
	\bibliographystyle{apsrev4-1}
	
\end{document}